\documentclass[letterpaper, 10 pt, journal, twoside]{ieeetran}

\usepackage[hidelinks]{hyperref}

\usepackage{mathrsfs}
\usepackage{bbold}
\usepackage{graphicx,amstext,amsmath,amssymb,color,psfrag}
\usepackage[latin1]{inputenc}

\newcommand{\Rlo}{\ensuremath{\mathbb{R}_{\geq 0}}}

\definecolor{bleucit}{rgb}{0.2,0.4,0.6} 

\definecolor{blue_cv}{rgb}{0.09,0.35,0.78}

\newcommand{\Argmin}{\ensuremath{\text{argmin}\,}}
\DeclareMathOperator*{\argmin}{\Argmin}

\newcommand{\Argmax}{\ensuremath{\text{argmax}\,}}
\DeclareMathOperator*{\argmax}{\Argmax}

\newtheorem{ass}{Assumption}

\newtheorem{prop}{Proposition}

\newtheorem{thm}{Theorem}

\newtheorem{rem}{Remark}

\usepackage{diagbox}
\usepackage{caption}

\usepackage{subcaption}

\title{A Stackelberg viral marketing design for two competing players$^*$\thanks{$^*$This work  was supported by ANR through the grant NICETWEET No. ANR-20-CE48-0009 and the PNRR project DECIDE No. 760069.}}

\author{ O. Lindamulage De Silva$^1$\thanks{$^1$ Universit\'e de Lorraine, CNRS, CRAN, F-54000 Nancy, France, {\scriptsize \tt olivier.lindamulage-de-silva@univ-lorraine.fr}}, V. S. Varma$^{1}$, M. Cao$^2$\thanks{$^2$ Faculty of Mathematics and Natural Sciences, ITM, University of Groningen, The Netherlands}, I.-C. Mor\u{a}rescu$^{1,3}$\thanks{$^3$associated with Automation Department, Technical University of Cluj-Napoca, Memorandumului 28, 400114 Cluj-Napoca, Romania}, S. Lasaulce $^1$\\}

\begin{document}
\maketitle
\thispagestyle{empty}

\begin{abstract}
A Stackelberg duopoly model in which two firms compete to maximize their market share is considered. The firms offer a service/product to customers that are spread over several geographical regions (e.g., countries, provinces, or states). 
Each region has its own characteristics (spreading and recovery rates) of each service propagation. We consider that the spreading rate can be controlled by each firm and is subject to some investment that the firm does in each region. 
One of the main objectives of this work is to characterize the advertising budget allocation strategy for each firm across regions to maximize its market share when competing. To achieve this goal we propose a Stackelberg game model that is relatively simple while capturing the main effects of the competition for market share. {By characterizing the strong/weak Stackelberg equilibria of the game, we provide the associated budget allocation strategy.} In this setting, it is established under which conditions the solution of the game is the so-called ``winner takes all". Numerical results expand upon our theoretical findings and we provide the equilibrium characterization for an example. 
\end{abstract}
 
  \section{Introduction}

Viral marketing (VM) is a strategy to promote different products/services using social networks. The information spreads as a virus from one person to {their} family, friends and colleagues. { To model the spread of information or services within a population, various mathematical frameworks have been developed. In the context of viral marketing, two prominent modeling approaches are opinion dynamics models (see \cite{Tu-ACMWeb2022,moruarescu2020space} for instance) and epidemic models (see \cite{liu2017continuous,liu2019analysis} for instance). Opinion dynamics models focus on capturing how opinions and preferences evolve among individuals in a social network, considering factors such as interaction pattens, biased influences, and coupled decision-making processes. On the other hand, epidemic models capture the transmission of diseases or information within a population, incorporating parameters such as infection rates, recovery rates, and population connectivity, which play a crucial role in understanding the dynamics of service propagation and devising effective marketing strategies to maximize the spread and adoption of the service. In this paper, we are not focusing on a specific dynamic but assume that the equilibrium in each region corresponds to the so-called ``winner takes all" \cite{Srinivasan2021}, for which a particular case of such model is the steady-state of a multi-virus SIS system \cite{prakash2012winner}. It is noteworthy that such an equilibrium has been observed in real life, see for instance the classical example of Facebook and Myspace \cite{prakash2012winner}. In the context of viral marketing duopoly, the concept of ``winner takes all" refers to a scenario where two competing firms engage in viral marketing campaigns to gain a larger market share. In this situation, the firm that effectively leverages viral marketing techniques and achieves widespread adoption of its product or service tends to capture the majority of the market, leaving the competitor with a significantly smaller share.}

It has been proven that targeted marketing combined with social network spreading has advantages over conventional mass campaigns, including cost-effectiveness and the ability to reach specific customer groups \cite{moruarescu2020space}. Basically, the authors formulated the problem as an optimal budget allocation and they shown that most individuals have to be targeted by the marketing campaign in order to get a maximum profit. The problem of competition to get a larger market share has been addressed in \cite{varma2017opinion}. The authors introduced a duopoly model which accounts for the knowledge of opinion dynamics through a social network and characterized the Nash strategies of the players.  
Unlike these works that take advantage of the node centrality (network topology) but rely on linear opinion dynamic models for the service spreading, we assume here {a winner takes all model based on \cite{prakash2012winner}.} This setup is more suitable for certain types of products/services, such as video streaming and activities in social networking platforms. 

The main goal of this paper is to formulate a two-level Stackelberg game and characterize its solutions. The analysis does not rely on a specific dynamic for the information spreading but assumes that at the equilibrium the ``winner takes all". We point out that the existing literature on multi-virus epidemic emphasize the aforementioned type of equilibrium \cite{prakash2012winner,liu2017continuous,liu2019analysis}. We consider that two players compete over several regions to get a higher market share. The budget allocated by a firm to a certain region modifies the spreading rate of the associated service in that region. 
Note that we are analyzing the steady state revenue of each firm. This combined with the ``winner takes all" behavior allows us to decouple the analysis of the investment strategy for each region. At the same time, we highlight that the design of the budget allocation strategy is subject to overall fixed budget constraints.

Note that each player has to solve a budget allocation problem which is different from the ones that can be found in the literature. In \cite{gracy2022analysis}, a dynamic optimization problem under budget constraints is formulated to control a single-virus SIS model. In \cite{taynitskiy2020optimal}, optimal control of joint multi-virus infection and information dissemination is considered for the sensitive-warned-infected-recovered-susceptible (SWIRS) model, without budget constraints. The differences between the present work and the existing results on epidemic control (e.g., \cite{LCSSO,kandhway2014run, nowzari2015optimal,zaric2002dynamic,preciado2013optimal}) are mainly related to the fact that our model {can} handle a multi-virus SIS epidemic model by considering budget constraints. Two existing works are relatively close to the present one. The first is \cite{varma2022non} in which the authors formulate a static and strategic form game to deal with a bi-virus SIS epidemic model over a single region without a budget constraint. The second one is \cite{OlivierECC2023} in which only one player solves the optimal budget allocation problem over several regions.

This paper is structured as follows. {Sec. II provides the problem formulation, going from the VM model 
up to the Stackelberg game analyzed in the paper.} The main result of this work is presented at the end of Sec. II. Sec. III is devoted to the proof of the main result. A numerical example illustrates our theoretical findings in Sec. IV and some concluding remarks are provided in Sec. V.

\section{Problem statement}\label{sec:SingleRegion}

We consider a set of two firms competing over $K > 1$ regions (e.g., countries, provinces, or states) to maximize their market share. Let $\mathcal{K}:=\{1,\ldots,K\}$ and $\mathcal{M}:=\{1,2\}$ be the set of regions and firms, respectively. For a given region $k \in \mathcal{K}$ and a firm $m\in\mathcal{M}$, we respectively denote by (i) $\gamma_{mk}$ the spreading rate of the service of Firm $m$ in the region $k$; (ii) $\delta_{mk}$ the churn rate at which the individuals from Region $k$ decide to dispense with the services of Firm $m$. {In this work, we assume that there is a simple linear relationship between allocated budget and $\gamma_{mk}$ dissemination rate, but more advanced and faithful models could be investigated in future work. In practice, the spread rate is influenced by numerous factors and it would be more accurate to assume that firms focus on captivating content and identifying influential users or communities to indirectly influence the spread rate of viral marketing campaigns \cite{rodrigues2016can,KANDHWAY201479,rodrigues2015viral}.} On top of this, we assume that Firm $m$ has a given advertising budget $B_m$ to allocate between the $K$ regions in order to maximize the number of its subscribers. We denote by $\gamma_m:=(\gamma_{m,1},\ldots,\gamma_{mk})\in\Rlo^{K}$ the action vector of Firm $m$ {such that $\sum_{k=1}^K\gamma_{mk}\leq B_m$} and $\gamma:=(\gamma_1,\gamma_2) \in\Rlo^{2K}$ the  action profile. We also make a slight notation abuse by using the following notation: $\gamma:=(\gamma_m,\gamma_{-m})$ for $m\in\mathcal{M}$.

Firm $1$ will be referred to as the \emph{leader} and Firm $2$ as the \emph{follower}. This is because the leader acts on the network first and the follower acts after a sufficiently long time such that it can observe and react to the action made by the leader.


\vspace{-1em}
\subsection{Viral marketing model} \label{subsec:epidemic-model1} 
By considering that the control action $\gamma_{mk}$ is a constant on the working phase $[0,+\infty[$. Thus, the follower can be said to start influencing {nodes} at time $0$. The fraction of individuals in Region $k\in\mathcal{K}$ who subscribe to the services of Firm $m\in\mathcal{M}$ is denoted by $x_{mk}\in[0,1]$. {For the interest of practicality, we consider the following assumption.
\begin{ass}
Each individual subscribes to at most one service.
\end{ass}This assumption is justified by the consideration of average individual behavior, which aligns with the notion that individuals typically opt for a single subscription.} The fraction of individuals in Region $k\in\mathcal{K}$ who have not subscribed to any services is denoted by $s_k\in[0,1]$. In what follows we denote by $x_{mk}^{\infty}:=\lim\limits_{t\to\infty}x_{mk}(t)$  $\forall k\in\mathcal{K},\forall m\in\mathcal{M}$ and we suppose that:
\begin{equation}\label{eq:steadystate}
    \hspace{-0.5em}(x_{1k}^\infty,x_{2k}^\infty)=\left\{\begin{array}{ll}
        \displaystyle \Big(1-\frac{\delta_{1k}}{\gamma_{1k}},0\Big)& \text{if }\displaystyle\frac{\gamma_{1k}}{\delta_{1k}}\geq \max\left(\displaystyle\frac{\gamma_{2k}}{\delta_{2k}},1\right)\\
           \displaystyle \Big(0,1-\frac{\delta_{2k}}{\gamma_{2k}}\Big)& \text{if }\displaystyle\frac{\gamma_{2k}}{\delta_{2k}}\geq \max\left(\displaystyle\frac{\gamma_{1k}}{\delta_{1k}}+ \pi,1\right)\\
           (0,0)& \text{otherwise.}
    \end{array}\right.
\end{equation}
The presence of $\pi>0$ in \eqref{eq:steadystate} is motivated by practical reasons well-known in the economics literature. The leader has already established its market by the time when the follower enters the market. Thus, the follower has to invest a little more to convince the customers to switch services. In the economics literature $\pi$ is called {a barrier to entry. A barrier to entry is a condition that makes it difficult for new firms to enter a market and compete with established firms. Barriers can take various forms, such as economies of scale, brand loyalty, access to distribution channels, patents and copyrights, government regulations, and high capital requirements \cite{DemsetzHarold,Kbasu}. The analysis in \cite[Section 4.2]{prakash2012winner} and \cite{doshi2021competing} provide a bi-virus (Susceptible-Infected-Susceptible) SIS model applied for viral marketing, such that the system converges to a point closed to \eqref{eq:steadystate}. The main difference is that the case when $\displaystyle\frac{\gamma_{1k}}{\delta_{1k}}=\displaystyle\frac{\gamma_{2k}}{\delta_{2k}}$ leads to multiple equilibria, and is avoided in our model due to the barrier to entry $\pi$. Additionally, by setting the churn rates $\delta_{mk}$ to zero, the model corresponds to a Colonel Blotto game, which is commonly employed in viral marketing literature \cite{niknami2022competitive}.}

\vspace{-1em}
\subsection{Game model}\label{subsec:game-formulation}

As previously stated, each Firm $m \in\mathcal{M}$ solves a budget allocation problem that maximizes a revenue under the global budget (denoted by ${B}_m$) constraint i.e., a feasible strategy for Firm $m$ belong to the set $\Gamma_m:=\{\gamma_m\in\Rlo^K:\ \sum_{k=1}^K\gamma_{mk}\leq \mathrm{B_m}\}$. The budget ${B}_m$ is imposed for practical reasons (e.g., a firm has a given finite investment budget). 

Similar to \cite{xu2015competition}, we consider the utility of each Firm $m\in\mathcal{M}$, such as given by \begin{equation}
    \label{eq:initialutility}
    \resizebox{0.8\linewidth}{!}{$\begin{array}{l}
\displaystyle u_{1}(\gamma_1,\gamma_{2}):=\sum_{k=1}^K p_{1k}\left(1-\frac{\delta_{1k}}{\gamma_{1k}}\right)\resizebox{0.34\linewidth}{!}{$\mathbb{1}_{\mathcal{H}_1(k,\gamma_{2k})}(\gamma_{1k})$}\\
  \displaystyle u_{2,\pi}(\gamma_2,\gamma_{1}):=\sum_{k=1}^K p_{2k}\left(1-\frac{\delta_{2k}}{\gamma_{2k}}\right)\resizebox{0.34\linewidth}{!}{$\mathbb{1}_{\mathcal{H}_{2,\pi}(k,\gamma_{1k})}(\gamma_{2k})$}
    \end{array}$}
\end{equation}
where: (i) $\pi>0$ is {the barrier to entry} for the follower, which is fixed; (ii) $p_{mk}$ the contribution of Region $k$ to the revenue of firm $m$ when all its individuals subscribe to the service $m$; (iii) the quantity $ \big(1-\frac{\delta_{mk}}{\gamma_{mk}}\big)$ represents $x_{mk}^\infty(\gamma,\pi)$ in view of \eqref{eq:steadystate} and $\mathbb{1}$ is the indicator function, where
\[\mathcal{H}_1(k,\gamma_{2k}):=\left\{\gamma_{1k}\in\Rlo:\  \displaystyle\frac{\gamma_{1k}}{\delta_{1k}}\geq \max\left( \displaystyle\frac{\gamma_{2k}}{\delta_{2k}},1\right)\right\},\]
and
\[\hspace{-0.5em}\mathcal{H}_{2,\pi}(k,\gamma_{1k}):=\left\{\gamma_{2k}\in\Rlo:\  \displaystyle\frac{\gamma_{2k}}{\delta_{2k}}\geq \max\left( \displaystyle\frac{\gamma_{1k}}{\delta_{1k}}+\pi,1\right)\right\}.\]

{In this paper, we analyze a static Stackelberg game with a leader (Firm 1) and a follower (Firm 2). The utility of Firm 1 depends only on $\gamma_1$ as Firm 2 reacts with a best response strategy. We adopt a pessimistic approach for the leader, focusing on the analysis of weak Stackelberg equilibrium \cite{leitmann1978generalized}. The leader's utility function, considering a given barrier to entry $\pi>0$ for the follower, is formulated as: 
\begin{equation} \label{eq:utilityStackelberg}
\left.\begin{array}{ll}
\displaystyle u_{1,\pi}^S(\gamma_1):=\min\limits_{\gamma_2\in\mathrm{BR}_{2,\pi}(\gamma_1)} u_1(\gamma_1,\gamma_2),\\
\displaystyle\mathrm{BR}_{2,\pi}(\gamma_1)=\argmax_{\gamma_2\in\Gamma_2}u_{2,\pi}(\gamma_2,\gamma_1).
\end{array}\right.
\end{equation}
It should be noted that the analysis presented in this paper also applies when considering an optimistic formulation for the leader, namely, strong Stackelberg equilibrium with $u_{1,\pi}^S(\gamma_1)=\max\limits_{\gamma_2\in\mathrm{BR}_{2,\pi}(\gamma_1)}u_1(\gamma_1,\gamma_2)$.} The goal of this work is to analyse {a regular Stackelberg solution (S)}, for a given $\pi>0$ of the
game $\mathcal{G}_{\pi}^S:=\left(\mathcal{M},(\Gamma_m)_{1\leq m\leq 2},(u_{m,\pi}^S)_{1\leq m \leq 2}\right)$ in which: the players are Firm $1$ and Firm $2$; the action space of Firm $m$ is given by $\Gamma_m$; the individual utility function of each firm is given by $u_{1,\pi}^S$ in (\ref{eq:utilityStackelberg}) and $u_{2,\pi}^S=u_{2,\pi}$ in \eqref{eq:initialutility}. Firm $m\in\mathcal{M}$ expresses its interests by setting the potential revenues $(p_{m,1},\ldots,p_{mk})\in\Rlo^K$, whereas the set of action $\Gamma_m$ is imposed by the capacity of investment of Firm $m$ given by the budget $\mathrm{B_m}$. In addition, we emphasize that the theoretical results established in this paper hold for a multistage game setup in which the one-shot game is repeated at each stage (for which the parameters are updated) and different constant control actions are applied during it.

Let us recall the definition of a {weak} Stackelberg solution. The strategy ${\gamma}_{\pi}^{\mathrm{S}}$  is {a weak Stackelberg solution} of the game $\mathcal{G}_{\pi}^S$ if it is a solution of the system of equations: 
\[\left.\begin{array}{ll}
\gamma_{1,\pi}^{\mathrm{S}}\in\argmax\limits_{\gamma_1\in\Gamma_1}u_{1,\pi}^S(\gamma_1)\text{ and }
     \gamma_{2,\pi}^{\mathrm{S}}\in\argmax\limits_{\gamma_2\in\Gamma_2} u_{2,\pi}^S(\gamma_2,\gamma_{1,\pi}^{\mathrm{S}}).
\end{array}\right.\] 
\subsection{Main result}
 As mentioned above, we investigate the weak Stackelberg solution of the game $\mathcal{G}_{\pi}^S$. In what follows, we denote by:
 \[\mathrm{BR}_{2,\pi}^{-}(\gamma_1):=\argmin\limits_{\gamma_2\in\mathrm{BR}_{2,\pi}(\gamma_1)}u_1(\gamma_1,\gamma_2)\] the best response of Firm $2$ which minimizes the profit of the Firm $1$ when it uses the strategy $\gamma_1\in\Gamma_1$. Let $\mathrm{BR}_{2k,\pi}^{-}(\gamma_1):=\mathrm{P}_k(\mathrm{BR}_{2,\pi}^{-}(\gamma_1))$ where $\forall \gamma_2\in\Gamma_2,\ k\in\mathcal{K}$ one has $\mathrm{P}_k(\gamma_2)=\gamma_{2k}$.
 
 
{Let $\mathcal{K}_1\in 2^K$ representing a given set of regions of investment under consideration by Firm 1}, let us define the following set  \[\begin{array}{ll}
\widehat{\Gamma}_{1,\pi}(\mathcal{K}_1):=\Bigg\{\gamma_1\in\Rlo^K \Big\vert\\
\displaystyle \forall k\in\mathcal{K}_1,\ \frac{\gamma_{1k}}{\delta_{1k}}\geq\max\left(\frac{\mathrm{BR}_{2k,\pi}^{-}(\gamma_1)}{\delta_{2k}},1\right),\\
\displaystyle\forall k\in\mathcal{K}\setminus\mathcal{K}_1,\ \frac{\gamma_{1k}}{\delta_{1k}}\leq\max\left(\frac{\mathrm{BR}_{2k,\pi}^{-}(\gamma_1)}{\delta_{2k}}-\pi,1\right)\Bigg\},
\end{array}\] 
and $\widehat{S}_{\pi}(\mathcal{K}_1,\gamma_1):=$
\begin{equation}\label{eq:hatS}
\left\{\hspace{-0.5cm}\begin{array}{ll}
&\sum_{k\in\mathcal{K}_1}p_{1k}\left(\displaystyle 1-\frac{\delta_{1k}}{\gamma_{1k}}\right)\text{ if }\gamma_1\in\widehat{\Gamma}_{1,\pi}(\mathcal{K}_1)\neq\emptyset,\\
&-1\hspace{3.1cm}\text{ otherwise. }
 \end{array}\right.
\end{equation}
 {In simpler terms, the expression $\widehat{\Gamma}_{1,\pi}(\mathcal{K}_1)$ represents the set of investment options available to Firm 1 when Firm 2 applies a best response strategy. This set allows Firm 1 to determine the regions and values of investment that it can choose from. On the other hand, $\widehat{S}_{\pi}(\mathcal{K}_1,\gamma_1)$ represents the corresponding revenue that Firm 1 can generate based on its chosen investment strategy $\mathcal{K}_1$ and spread rate $\gamma_1$.} 
We are now ready to state the main result of this paper, which will be proven in the following section. 
\begin{thm}\label{thm:Thm1}
The Stackelberg strategy $\gamma_\pi^S$ is obtained by solving:
\begin{equation}\label{eq:StackSol1}
(\mathcal{K}_{1,\pi}^S,\gamma_{1,\pi}^S):=\argmax_{\mathcal{K}_1\in 2^\mathcal{K}}\argmax_{\gamma_1\in\Gamma_1}\widehat{S}_{\pi}(\mathcal{K}_1,\gamma_1)
\end{equation}
where $\widehat \Gamma_{1,\pi}(\mathcal{K})$ {is proven to be a convex set (see proof)} for any $\mathcal{K}$ {and the maximization problem $\max\limits_{\gamma_1\in\Gamma_1}\widehat{S}_\pi(\mathcal{K}_1,\gamma_1)$ is a convex OP for a given $\mathcal{K}_1$.} The follower's strategy at the Stackelberg equilibria is given by $\gamma_{2k,\pi}^{S}=\frac{\displaystyle\sqrt{p_{2k}\delta_{2k}}\displaystyle}{\displaystyle\sum_{\ell\in{\mathcal{K}}_{2,\pi}^{S}}\sqrt{p_{2\ell}\delta_{2\ell}}}\mathrm{B_2}$ if $k\in{\mathcal{K}}_{2,\pi}^{S}$ and $\gamma_{2k,\pi}^{S}=0$ otherwise, where
\[\begin{array}{l}
\displaystyle\mathcal{K}_{2,\pi}^S:=\argmax_{\mathcal{K}_2\in 2^\mathcal{K}}\sum_{k\in\mathcal{K}_2}p_{2k}\left(1-\frac{\displaystyle\sqrt{\frac{\delta_{2k}}{p_{2k}}}\sum_{\ell\in\mathcal{K}_2}\sqrt{p_{2\ell}\delta_{2\ell}}}{\mathrm{B_2}}\right)\\
\hspace{2.5cm}\text{s.t}.\ \mathcal{K}_2\cap\mathcal{K}_{1,\pi}^S=\emptyset.
\end{array}\]
\end{thm}

\begin{rem}
In the perspective of studying the {strong} Stackelberg solution, it is enough to consider that the follower applies the strategy in $\mathrm{BR}_{2,\pi}^{+}(\gamma_1)$ where
\[\mathrm{BR}_{2,\pi}^{+}(\gamma_1):=\argmax\limits_{\gamma_2\in\mathrm{BR}_{2,\pi}(\gamma_1)}u_1(\gamma_1,\gamma_2).\]
\end{rem}
\begin{rem}{
Although the optimization problem specified in Equation \eqref{eq:StackSol1} theoretically involves an exponential search space of $2^K$ regions. On one hand we provide just a methodology to solve such problems (that does not involve online solving and therefore computation time is not very important) and on the other hand, in many practical scenarios, the number of regions $K$ in which firms compete is relatively small, typically around $10$. Basically we can consider that USA, Europe, South America represent regions for competition at global level. If the competition is specific to a country than the number of regions will be also relatively small. This manageable number of regions significantly reduces the computational complexity associated with solving the optimization problem. A key result introduced in Theorem \ref{thm:Thm1} is by ensuring that the maximization of $\widehat{S}_\pi(\mathcal{K}_1,\gamma_1)$, for a given $\mathcal{K}_1$ and for $\gamma_1\in\Gamma_1$ can be effectively resolved through the utilization of a convex optimization algorithm.}
\end{rem}
In order to prove the main result of this paper, we first characterize the best response strategy of Firm $2$ and finally characterize the weak Stackelberg solution for a fixed {barrier to entry} $\pi>0$.
\section{Stackelberg strategy design}
 In this section, we propose a convex reformulation of the problem in \eqref{eq:utilityStackelberg}. This allows us to obtain the optimal budget allocation solution for the follower and thus characterize the weak Stackelberg solution of the game $\mathcal{G}_{\pi}^S$.

For a given leader's strategy $\gamma_{1}\in\Gamma_1$ we denote by $\widetilde{\Gamma}_{2,\pi}(\mathcal{K}_2,\gamma_{1})$ as the set of strategy of Firm $2$ to win the marketing battles in regions of $\mathcal{K}_2\subseteq \mathcal{K}$ i.e.,\\ $\widetilde{\Gamma}_{2,\pi}(\mathcal{K}_2,\gamma_{1}):=\Bigg\{\gamma_2\in \Rlo^K\Bigg\vert$
\begin{equation}\label{eq:widetildeGammam}
\begin{array}{ll}
\displaystyle \forall k\in\mathcal{K}_2,\ \frac{\gamma_{2k}}{\delta_{2k}}\geq\max\left(\frac{\gamma_{1k}}{\delta_{1k}}+\pi,1\right),\\
\displaystyle\forall k\in\mathcal{K}\setminus\mathcal{K}_2,\ \frac{\gamma_{2k}}{\delta_{2k}}\leq\max\left(\frac{\gamma_{1k}}{\delta_{1k}},1\right)\Bigg\}.
\end{array}
\end{equation}


\subsection{Follower's OP reformulation}
The goal of this section is to reformulate the optimisation problem for the follower ($m=2$) introduced in \eqref{eq:utilityStackelberg}. 
\begin{prop}\label{prop:prop1}
Let $\gamma_{1}\in\Gamma_{1}$. The initial OP in \eqref{eq:utilityStackelberg} can be reformulated such as $ \displaystyle\max_{\gamma_2\in\Gamma_2}u_{2,\pi}^S(\gamma_2,\gamma_1)=$
\[\displaystyle \max_{\mathcal{K}_2\in 2^\mathcal{K}}\left[\begin{array}{ll}
     &\displaystyle \max_{\gamma_2}\displaystyle\sum_{k\in\mathcal{K}_2}p_{2k}\left(1-\frac{\delta_{2k}}{\gamma_{2k}}\right)\\
&\hspace{1cm}\text{s.t.}\displaystyle\sum_{k=1}^K\gamma_{2k}\leq \mathrm{B_2}\\
&\hspace{1.5cm}\gamma_2\in\widetilde{\Gamma}_{2,\pi}(\mathcal{K}_2,\gamma_1).
\end{array}\right]\]
\[\label{eq:ReformulatedPb}\tag{$P^\star$}
\hspace{1.7cm}\text{s.t. }\exists\gamma_2\in\widetilde{\Gamma}_{2,\pi}(\mathcal{K}_2,\gamma_1):\ \sum_{k=1}^K\gamma_{2k}\leq \mathrm{B_2}\]
\end{prop}
\begin{rem}\label{RemarkconditionKm}
It appears that, the condition $``\exists \gamma_2\in\widetilde{\Gamma}_{2,\pi}(\mathcal{K}_2,\gamma_{1}):\ \sum_{k=1}^K\gamma_{2k}\leq\mathrm{B_2}''$ is verified, if and only if  $\displaystyle\sum_{k\in\mathcal{K}_2}\delta_{2k}\max\left(\displaystyle\frac{\gamma_{1k}}{\delta_{1k}}+\pi,1\right)\leq \mathrm{B_2}$ i.e., the less restrictive strategy of $\widetilde{\Gamma}_{2,\pi}(\mathcal{K}_2,\gamma_1)$ verifies the budget constraint.
\end{rem}
\noindent\emph{$\mathbf{Proof.}$}
In view of \eqref{eq:utilityStackelberg}, the follower's OP can be rewritten in the following manner by considering both cases of Firm $2$ winning or not in region $k$:
\[\label{eq:P_1} \tag{$P_1$}\displaystyle\max_{\gamma_2\in\Gamma_2}\sum_{\mathcal{K}_2\in 2^\mathcal{K}}\left[\sum_{k\in\mathcal{K}_2}p_{2k}\left(1-\frac{\delta_{2k}}{\gamma_{2k}}\right)\right]\mathbb{1}_{\widetilde{\Gamma}_{2,\pi}(\mathcal{K}_2,\gamma_{1})}(\gamma_2),\]
where $\widetilde{\Gamma}_{2,\pi}$ is introduced in \eqref{eq:widetildeGammam}  and $\mathbb{1}$ is the indicator function.

Let $\gamma_2,\widehat{\gamma}_2\in\Gamma_2$, $\mathcal{K}_2,\widehat{\mathcal{K}}_2\in 2^\mathcal{K}$ such that $\mathcal{K}_2\neq \widehat{\mathcal{K}}_2$ and $\mathbb{1}_{\widetilde{\Gamma}_{2,\pi}(\mathcal{K}_2,\gamma_{1})}(\gamma_2)=\mathbb{1}_{\widetilde{\Gamma}_{2,\pi}(\widehat{\mathcal{K}}_2,\gamma_{1})}(\widehat{\gamma}_2)=1$. Then,  $\widetilde{\Gamma}_{2,\pi}(\mathcal{K}_2,\gamma_{1})\cap\widetilde{\Gamma}_{2,\pi}(\widehat{\mathcal{K}}_2,\gamma_{1})=\emptyset$ since $\mathcal{K}_2\neq \widehat{\mathcal{K}}_2$ and \eqref{eq:P_1} can be reformulated such as 
\[\max\limits_{\mathcal{K}_2\in 2^\mathcal{K}}\Bigg[\displaystyle\max_{\gamma_2\in\Gamma_2}\Big[\sum_{k\in\mathcal{K}_2}p_{2k}\left(1-\frac{\delta_{2k}}{\gamma_{2k}}\right)\Big]\mathbb{1}_{\widetilde{\Gamma}_{2,\pi}(\mathcal{K}_2,\gamma_{1})}(\gamma_2)\Bigg].\label{eq:P_2}\tag{$P_2$}\]
Furthermore, for a given $\mathcal{K}_2$ there exits a strategy $\gamma_2\in\widetilde{\Gamma}_{m}(\mathcal{K}_2,\gamma_{1})$ that verifies the budget constraint if and only if the set $\mathcal{K}_2$ verifies $
\displaystyle\sum_{k\in\mathcal{K}_2}\delta_{2k}\max\left(\displaystyle\frac{\gamma_{1k}}{\delta_{1k}}+\pi,1\right)\leq \mathrm{B_2}$ i.e., the less restrictive action of $\widetilde{\Gamma}_{2,\pi}(\mathcal{K}_2,\gamma_{1})$ verifies the budget constraint. Hence, by using the indicator function and by adding the new constraint on the feasibility sets of $\mathcal{K}_2$, the problem \eqref{eq:P_2} is equivalent to \eqref{eq:ReformulatedPb}.\hfill$\blacksquare$\vspace{-0.2cm} 

\subsection{Characterization of the follower's best response}

The following proposition establishes the characterization of the best response $\mathrm{BR}_{2,\pi}$ in \eqref{eq:utilityStackelberg} for the follower and for a given strategy $\gamma_{1}$ of the leader.
\begin{prop}\label{prop:prop2}

 Let $\gamma_{1}\in\Gamma_{1}$. The  Best response of the follower is characterized by:\hfill
\[\begin{array}{l}
(\mathcal{K}_{2,\pi}^{\mathrm{BR}}(\gamma_1),\widetilde{\mathcal{K}}_{2,\pi}^{\mathrm{BR}}(\gamma_1))\in\\
\displaystyle\displaystyle\argmax\limits_{\mathcal{K}_2,\widetilde{\mathcal{K}}_2}\sum\limits_{k\in\mathcal{K}_2\setminus\widetilde{\mathcal{K}}_2}p_{2k}\left(1-\frac{1}{\displaystyle\max\left(\displaystyle\frac{\gamma_{1k}}{\delta_{1k}}+\pi,1\right)}\right)\\
\displaystyle+\sum_{k\in\mathcal{K}_2}p_{2k}\left(1-\displaystyle\frac{\displaystyle\sqrt{\frac{\delta_{2k}}{p_{2k}}}\sum_{\ell\in\widetilde{\mathcal{K}}_2}\sqrt{p_{2\ell}\delta_{2\ell}}}{\displaystyle\mathrm{B_2}-\sum_{\ell\in\mathcal{K}_2\setminus\widetilde{\mathcal{K}}_2}\delta_{2\ell}\max\left(\displaystyle\frac{\gamma_{1\ell}}{\delta_{1\ell}}+\pi,1\right)}\right)\\
\text{s.t. }\forall k\in\widetilde{\mathcal{K}}_2,\\
\frac{\displaystyle\sqrt{p_{2k}\delta_{2k}}\left[\displaystyle\mathrm{B_2}-\sum_{\ell\in\mathcal{K}_2\setminus\widetilde{\mathcal{K}}_2}\delta_{2\ell}\max\left(\displaystyle\frac{\gamma_{1\ell}}{\delta_{1\ell}}+\pi,1\right)\right]}{\displaystyle\sum_{\ell\in\widetilde{\mathcal{K}}_2}\sqrt{p_{2\ell}\delta_{2\ell}}}\\
\hspace{4cm}>\delta_{2k}\max\left(\displaystyle\frac{\gamma_{1k}}{\delta_{1k}}+\pi,1\right).
\end{array}\]

Furthermore, any $\gamma_{2,\pi}^\mathrm{BR}\in\mathrm{BR}_{2,\pi}(\gamma_1)$ is defined by:
$\gamma_{2k,\pi}^{\mathrm{BR}}=$
\begin{equation}\label{eq:BRcharact}\hspace{-0.2em}\left\{\begin{array}{ll}
0\hspace{3.6cm}\mbox{ if } k\in\mathcal{K}\setminus{\mathcal{K}_{2,\pi}^{\mathrm{BR}}(\gamma_1)},\\
\delta_{2k}\max\left(\displaystyle\frac{\gamma_{1k}}{\delta_{1k}}+\pi,1\right) \hspace{0.5cm}\mbox{ if } k\in\mathcal{K}_{2,\pi}^{\mathrm{BR}}(\gamma_1)\setminus\widetilde{\mathcal{K}}_{2,\pi}^{\mathrm{BR}}(\gamma_1),\\
\hspace{-0.7em}\frac{\displaystyle\sqrt{p_{2k}\delta_{2k}}\left[\displaystyle\mathrm{B_2}-\hspace{-0.7cm}\sum_{\ell\in\mathcal{K}_{2,\pi}^{\mathrm{BR}}(\gamma_1)\setminus\widetilde{\mathcal{K}}_{2,\pi}^{\mathrm{BR}}(\gamma_1)}\hspace{-0.5cm}\gamma_{2\ell}^{\mathrm{BR}}\right]}{\displaystyle\sum_{\ell\in\widetilde{\mathcal{K}}_{2,\pi}^{\mathrm{BR}}(\gamma_1)}\sqrt{p_{2\ell}\delta_{2\ell}}},\ \mbox{ if } k\in\widetilde{\mathcal{K}}_{2,\pi}^{\mathrm{BR}}(\gamma_1).\\
\end{array}\right.
\end{equation}
\end{prop}
\noindent\emph{$\mathbf{Proof.}$} According to Proposition \ref{prop:prop1}, the best response of Firm $2$ is characterized by the best part $\mathcal{K}_2\in 2^\mathcal{K}$ that maximises \begin{equation}\label{eq:S2}
    \begin{array}{ll}
     &\displaystyle \max_{\gamma_2}\displaystyle\sum_{k\in\mathcal{K}_2}p_{2k}\left(1-\frac{\delta_{2k}}{\gamma_{2k}}\right)\\
&\text{s.t. }\gamma_2\in\widetilde{\Gamma}_{2,\pi}(\mathcal{K}_2,\gamma_1),\ \displaystyle\sum_{k=1}^K\gamma_{2k}\leq \mathrm{B_2}.
\end{array}
\end{equation}
Let us exploit the KKT conditions by defining first the Lagrangian: $\displaystyle\mathcal{L}(\gamma_2,\mu_2,\underline{\mu}_{2},\lambda_2):=\displaystyle\sum_{k\in\mathcal{K}\setminus\mathcal{K}_2}\underline{\mu}_{2k}\gamma_{2k}$
\[
\begin{array}{l}
+\displaystyle\sum\limits_{k\in\mathcal{K}_2}p_{2k}\left(1-\frac{\delta_{2k}}{\gamma_{2k}}\right) -\lambda_2\left(\sum_{k=1}^K\gamma_{2k}-\mathrm{B_2}\right)\\
\displaystyle+\sum_{k\in{\mathcal{K}}_2}\mu_{2k}\Bigg(\gamma_{2k}-\delta_{2k}\max\left(\displaystyle\frac{\gamma_{1k}}{\delta_{1k}}+\pi,1\right)\Bigg)\\
\displaystyle-\sum_{k\in\mathcal{K}\setminus{\mathcal{K}_2}}\mu_{2k}\Bigg(\gamma_{2k}\displaystyle-\delta_{2k}\max\left(\displaystyle\displaystyle\frac{\gamma_{1k}}{\delta_{1k}},1\right)\Bigg).\\
\displaystyle
\end{array}\] 
Let us denote by $\gamma_{2k,\pi}^{\star}$, $\mu_{2k,\pi}^{\star}$, $\lambda_{2,\pi}^{\star}$ and $\underline{\mu}_{2k,\pi}^{\star}$ the variables that verify the first-order optimality condition. 

For all $k\in{\mathcal{K}}\setminus\mathcal{K}_2$, $\displaystyle\frac{\partial \mathcal{L}}{\partial \gamma_{2k}}=-\lambda_{2,\pi}^{\star}-\mu_{2k,\pi}^{\star}+\underline{\mu}_{2k,\pi}^{\star}=0,$ then $\mu_{2k,\pi}^{\star}=0,\ \underline{\mu}_{2k,\pi}^{\star}=\lambda_{2,\pi}^{\star}>0\text{ and }\gamma_{2k,\pi}^{\star}=0.$

For all $k\in\mathcal{K}_2$, $\displaystyle\frac{\partial \mathcal{L}}{\partial \gamma_{2k}}=\displaystyle\frac{p_{2k} \delta_{2k}}{({\gamma}_{2k,\pi}^{\star})^2}-\lambda_{2,\pi}^{\star}-\mu_{2k,\pi}^{\star}=0$, then ${\gamma}_{2k,\pi}^{\star}=\displaystyle\sqrt{\frac{p_{2k}\delta_{2k}}{(\lambda_{2,\pi}^{\star}+\mu_{2k,\pi}^{\star})}}.$ Let $\widetilde{\mathcal{K}}_2\in 2^{\mathcal{K}_2}$ (that may be empty) such that, $\widetilde{\mathcal{K}}_2\in\{\widetilde{\mathcal{K}}\in 2^{\mathcal{K}_2}:\ \forall k\in\widetilde{\mathcal{K}},\ \mu_{2k,\pi}^{\star}=0\}.$ For all $k\in\widetilde{\mathcal{K}}_2$, the first-order optimality condition is verified when, $\displaystyle \gamma_{2k,\pi}^{\star}=\sqrt{\frac{p_{2k}\delta_{2k}}{\lambda_{2,\pi}^{\star}}}
>\delta_{2k}\max\left(\displaystyle\frac{\gamma_{1k}}{\delta_{1k}}+\pi,1\right)$
 and
  $\gamma_{2k,\pi}^{\star}=\delta_{2k}\max\left(\displaystyle\frac{\gamma_{1k}}{\delta_{1k}}+\pi,1\right),\ \forall k\in\mathcal{K}_2\setminus\widetilde{\mathcal{K}}_2.$ 
  Since $\lambda_{2,\pi}^{\star}>0$, it follows that, $\sum_{k=1}^K\gamma_{2k,\pi}^{\star}=\mathrm{B_2}.$ 
  Hence, $\displaystyle\sum_{\ell\in\widetilde{\mathcal{K}}_2} \gamma_{2\ell,\pi}^{\star}=\mathrm{B_2}-\sum_{\ell\in\mathcal{K}_2\setminus\widetilde{\mathcal{K}}_2}\gamma_{2\ell,\pi}^{\star}$ $\Rightarrow$ $\sqrt{\lambda_{2,\pi}^{\star}}=\frac{\displaystyle\sum_{\ell\in\widetilde{\mathcal{K}}_2}\sqrt{p_{2\ell}\delta_{2\ell}}}{\displaystyle\mathrm{B_2}-\sum_{\ell\in\mathcal{K}_2\setminus\widetilde{\mathcal{K}}_2}\gamma_{2\ell}^{\star}}.$
Thus, the solution of \eqref{eq:S2} is characterized by
$\gamma_{2k,\pi}^{\star}=$
\[\hspace{-0.2em}\left\{\begin{array}{ll}
0\hspace{4cm}\mbox{ if } k\in\mathcal{K}\setminus{\mathcal{K}_2},\\
\delta_{2k}\max\left(\displaystyle\frac{\gamma_{1k}}{\delta_{1k}}+\pi,1\right)\hspace{0.85cm}\mbox{ if } k\in\mathcal{K}_2\setminus\widetilde{\mathcal{K}}_2,\\
\hspace{-0.7em}\frac{\displaystyle\sqrt{p_{2k}\delta_{2k}}\left[\displaystyle\mathrm{B_2}-\sum_{\ell\in\mathcal{K}_2\setminus\widetilde{\mathcal{K}}_2}\gamma_{2\ell,\pi}^{\star}\right]}{\displaystyle\sum_{\ell\in\widetilde{\mathcal{K}}_2}\sqrt{p_{2\ell}\delta_{2\ell}}}\
\mbox{ if } k\in\widetilde{\mathcal{K}}_2.\\
\end{array}\right.
\]
Finally, the best response of the follower is obtained by a selection of $\mathcal{K}_2$ and $\widetilde{\mathcal{K}}_2$ solving the discrete OP as stated in Proposition \ref{prop:prop2}. 
\hfill$\blacksquare$
\vspace*{-0.5em}
\subsection{Proof of Theorem \ref{thm:Thm1}}

In view of \eqref{eq:utilityStackelberg} and Proposition \ref{prop:prop1}, the utility of the leader can be reformulated such as

$\displaystyle u_{1,\pi}^S(\gamma_1)=\sum_{\mathcal{K}_1\in 2^\mathcal{K}}\Big[\sum_{k\in\mathcal{K}_1}p_{1k}(1-\frac{\delta_{1k}}{\gamma_{1k}})\mathbb{1}_{\widehat{\Gamma}_{1,\pi}(\mathcal{K}_1)}(\gamma_1)\Big]$, and we recall that $\widehat{\Gamma}_{1,\pi}(\mathcal{K}_1)$ is defined before Theorem \ref{thm:Thm1}.
Thus, we identify the same utility structure as in the proof of Proposition \ref{prop:prop1} and at the Stackelberg equilibria, it can be written as
\begin{equation}\label{eq:approxStackelberg}
u_{1,\pi}^S(\gamma_{1,\pi}^S)=
\max_{\mathcal{K}_1\in 2^\mathcal{K}}\max_{\gamma_1\in\Gamma_1}\widehat{S}_{\pi}(\mathcal{K}_1,\gamma_1)
\vspace*{-1em}
\end{equation}
where $\widehat{S}_{\pi}(\mathcal{K}_1,\gamma_1)$ is defined in \eqref{eq:hatS}. Concerning the existence of the Stackelberg strategy, the result is mainly based on the existence of a feasible solution in the set of constraints for the leader. Since $\mathcal{K}_1=\emptyset$ and $\gamma_1=(0,\ldots,0)$ is in the set of constraints, we derive that the game $\mathcal{G}_{\pi}^S$ has at least one Stackelberg equilibrium. 

In order to compute numerically \eqref{eq:approxStackelberg} with well known solvers for convex optimization problems, let us prove that $\widehat{\Gamma}_{1,\pi}(\mathcal{K}_1)$ is a convex set. Let $\gamma_1^a\in\widehat{\Gamma}_{1,\pi}(\mathcal{K}_1)$ and $\gamma_1^b\in\widehat{\Gamma}_{1,\pi}(\mathcal{K}_1)$. Hence $\forall k\in\mathcal{K}_1$ and $i\in\{a,b\}$,

$\hspace{1.7cm}\displaystyle\frac{\gamma_{1k}^i}{\delta_{1k}}\geq \max\left(\displaystyle\frac{\mathrm{BR}_{2k,\pi}^{-}(\gamma_1^i)}{\delta_{2k}},1\right).$

From the best response characterization of the follower in Proposition \ref{prop:prop2} one has that $\forall k\in\mathcal{K}_1$, $\mathrm{BR}_{2k,\pi}^{-}(\gamma_1^i)=0$. Hence $\forall k\in\mathcal{K}_1$, $\gamma_{1k}^i\geq \delta_{1k}$. Let $\tau\in(0,1)$ and $\gamma_{1k}^b\geq \tau \gamma_{1k}^a+(1-\tau)\gamma_{1k}^b\geq \gamma_{1k}^a.$ From the monotony of the best response of the follower w.r.t the action of the leader it follows that, $\forall k\in\mathcal{K}_1$

$\hspace{1.7cm}\mathrm{BR}_{2k,\pi}^{-}(\tau \gamma_{1k}^a+(1-\tau)\gamma_{1k}^b)=0.$\\
Hence, $\tau \gamma_{1k}^a+(1-\tau)\gamma_{1k}^b\in\widehat{\Gamma}_{1,\pi}(\mathcal{K}_1).$
Finally, \eqref{eq:approxStackelberg} is a strictly convex OP that can be solved with numerical solver for convex OP. Finally, at the Stackelberg equilibrium this analysis guarantees that $\forall k\in\mathcal{K}_{1,\pi}^S$ (region where the leader invest), $\mathrm{BR}_{2k,\pi}^{-}(\gamma_{1,\pi}^S)=0$, and in view of Proposition \ref{prop:prop2} we derive that the follower's strategy is given by $\gamma_{2k,\pi}^{S}=\frac{\displaystyle\sqrt{p_{2k}\delta_{2k}}\displaystyle}{\displaystyle\sum_{\ell\in{\mathcal{K}}_{2,\pi}^{S}}\sqrt{p_{2\ell}\delta_{2\ell}}}\mathrm{B_2}$ if $k\in{\mathcal{K}}_{2,\pi}^{S}$ and $\gamma_{2k,\pi}^{S}=0$ otherwise, where $\mathcal{K}_{2,\pi}^S$ is defined in Theorem \ref{thm:Thm1}.

\section{Numerical performance analysis}

 In this section, we illustrate the solutions of the Stackelberg game when both firms play pessimistically over a network of $K=5$ regions. The parameters of the game are given by: $p_{1}=(p_{11},\ldots,p_{15})=(1,2,3,4,5)$; $p_{2}=(p_{11},\ldots,p_{25})=(2,3,1,5,4)$; $\delta_{1}=(\delta_{11},\ldots,\delta_{15})=10^{-1}\times (5,4,3,2,1)$; $\delta_{2}=(\delta_{21},\ldots,\delta_{25})=10^{-1}\times (1,2,3,4,5)$ and the entry price for the follower is fixed at $\pi=10^{-6}$.
\begin{table}[h!]
\centering
\begin{tabular}{|l|c|c|c|c|}\hline
Budgets&$(\gamma_{11,\pi}^S,\gamma_{21,\pi}^S)$& $(\gamma_{12,\pi}^S,\gamma_{22,\pi}^S)$ & $(\gamma_{13,\pi}^S,\gamma_{23,\pi}^S)$ \\ \hline
  (0.6,0.6)& (0.2,0) & (0.4,0)& (0,0)   \\ \hline
(0.6,0.6)&  (0,0) & (0,0)& (0.6,0) \\ \hline
(0.6,5)&(0.6,0) & (0,1.298)& (0,1.377) \\ \hline
(5,0.6)& (0.5,0) & (1,0)& (1.5,0)  \\ \hline
(5,5)& (0.833,0) & (1.666,0)& (2.5,0)   \\ \hline
\end{tabular}
\end{table}
\begin{table}[h!]
\centering
\begin{tabular}{|l|c|c|}\hline
Budgets& $(\gamma_{14,\pi}^S,\gamma_{24,\pi}^S)$ & $(\gamma_{15,\pi}^S,\gamma_{25,\pi}^S)$ \\ \hline
  (0.6,0.6)&  (0,0.335)& (0,0.264)  \\ \hline
(0.6,0.6)&  (0,0.335)& (0,0.264)\\ \hline
(0.6,5)& (0,1.298)& (0,1.025)\\ \hline
(5,0.6)& (2,0)& (0,0.6)  \\ \hline
(5,5)& (0,2.79)& (0,2.20)  \\ \hline
\end{tabular}
\caption{\label{tab:allocation} Budget allocation at the Stackelberg Strategy for different value of $(\mathrm{B}_1,\mathrm{B}_2)$}
\end{table}

\begin{figure}[h]
 \centering
   \includegraphics[width=0.4\linewidth]{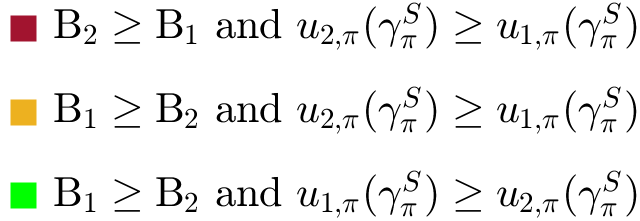}

 \includegraphics[width=\linewidth]{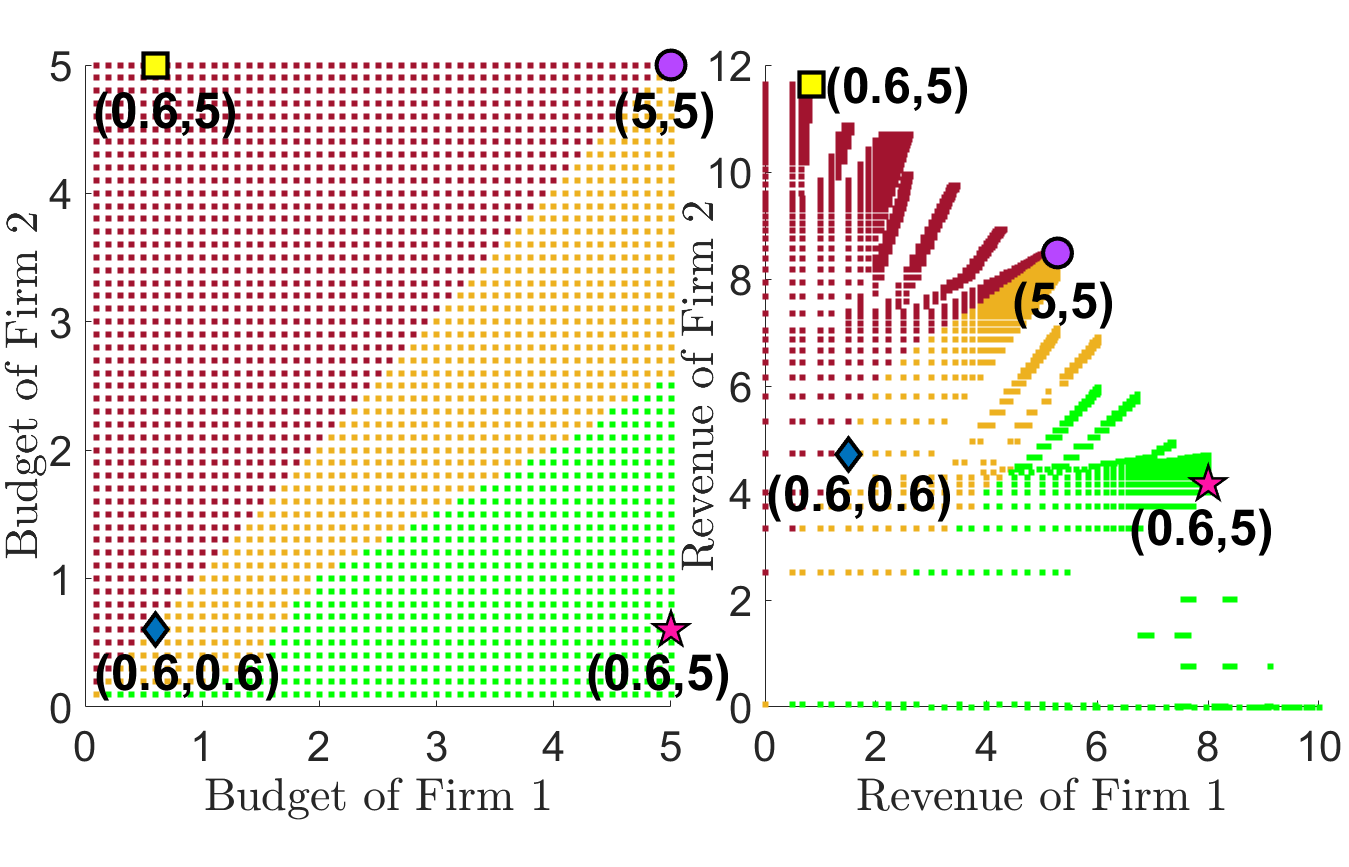}
     \caption{\label{fig:fig1} Revenue of each firm (right) at the Stackelberg equilbrium for values of $(\mathrm{B_1},\mathrm{B_2})$ shown on the left.}
     \vspace*{-0.5cm}
\end{figure}

Fig.~\ref{fig:fig1} represents the revenues of the two firms in the utility region, when they both apply Stackelberg's pessimistic strategy and for different budget values. In the case of equal budgets $\mathrm{B_1}=\mathrm{B_2}$ the follower has a higher revenue at the Stackelberg equilibrium due to the difference in the churn rate in regions 4 and 5. When one budget is much higher the corresponding Firm gets a larger revenue. We also observe that, at the Stackelberg equilibrium, there is no case where $\mathrm{B_1}\leq \mathrm{B}_2$ such that $u_{1,\pi}^S(\gamma_{\pi}^S)\geq u_{2,\pi}^S(\gamma_{\pi}^S)$. The TABLE.~\ref{tab:allocation} shows the Stackelberg strategy allocation for the couples $(\mathrm{B}_1,\mathrm{B}_2)$ highlighted in Fig.~\ref{fig:fig1}. 

\section{Conclusion}
We have formulated a Stackelberg duopoly game in which two firms compete for a larger market share when services propagate according a viral model. We have characterized analytically the corresponding Stackelberg strategy in the pessimistic/optimistic setting. To obtain this result, we have described the best response map of the follower and proved that the best response map of the leader can be found by solving a convex OP. 

To summarize, the paper provides a Stackelberg game to mathematically model a decision-making problem in a economic competition framework. The solution proposes strategic budget allocation across different regions in order to get a larger market share. It is noteworthy that we consider both the case of emerging companies (like startups) that need to cross a "barrier to entry" on the market and well established companies that want to preserve and enlarge their market share. Basically the paper gives insights on the strategic resource allocation with a priori given limitations in a competition setting.

\bibliographystyle{unsrt}
\bibliography{IEEEabrv,root}
\end{document}